\documentclass[12pt]{article}
\usepackage{epsfig}

\title{On a natural definition of the kilogram and the ampere
}
\author{Savely G. Karshenboim\\
\small \em D. I. Mendeleev Institute for Metrology (VNIIM), St.
Petersburg 190005, Russia\\
\small \em and Max-Planck-Institut f\"ur Quantenoptik, 85748
Garching, Germany\\
\small Email: sgk@vniim.ru, sek@mpq.mpg.de}
\begin{document}

\maketitle

\begin{abstract}
We consider a recent proposal \cite{mmqtw} to redefine the
kilogram in terms of natural constants. In our opinion, the main
objective of the redefinition should be to build such a version of
the SI system in which the electric measurements are possible with
the highest accuracy in SI units, and not in practical units as
now. We emphasize that this objective can be achieved only with a
simultaneous redefinition of the kilogram and ampere. This
redefinition must be in terms of fixed values of the Planck
constant $h$ and the elementary charge $e$. Certain details of the
possible redefinition are considered.
\end{abstract}

This paper considers the recent proposal \cite{mmqtw} to redefine
the SI kilogram and possibly the ampere in terms of fixed values
of fundamental physical constants. This would change the
International System of units (the SI) \cite{SI}, which is a
commonly accepted coherent system for all branches of macroscopic
measurements in education, sciences and technology. The
redefinition \cite{mmqtw}, which has been suggested in terms of
fundamental constants, also indirectly involves certain natural
quantum phenomena, which should appear due to realizations of the
redefinition.

Some time ago the SI was changed in a similar matter by fixing a
value of the speed of light $c$ \cite{c1983}. However, the present
situation is very different. To our opinion the major problem now
is that the present high-accuracy measurements in mechanics and
electricity are performed in two different versions of the SI.
While macroscopic mass measurements are performed in terms of the
SI kilogram, the most accurate electric measurements are performed
in terms of the practical units ohm-90 and volt-90. These latter
are apparently not consistent with the SI.

Microscopic mass measurements, however, are related to mass
determined in unified atomic mass units and in frequency units,
i.e., dealing with a value of $mc^2/h$ instead of the mass $m$.
Thus, they are measured in units closely related to the ohm-90 and
volt-90. For instance, the SI value of $h$ has a larger
uncertainty than microscopic mass comparisons, while in the
practical units-90 the numerical value of $h$ is known exactly.

The proposal \cite{mmqtw} and its numerous considerations in
various international commissions have been focussed on the
desirability of replacing the definition of the unit for mass
measurements, now based on the last artefact of the SI, the
kilogram prototype kept at the BIPM in S\`evres, by a definition
which is stable and independently reproducible. That mainly focuss
attention on the kilogram alone, while a redefinition of the
ampere is considered as one of a number of unnecessary collateral
options.

On contrary, we believe that the gap between the present version
of the SI and the system based on the ohm-90 and the volt-90 is a
crucial reason to consider such a redefinition. The modern version
of the SI \cite{SI} was introduced in 1983 by fixing the value of
the speed of light $c$ by CIPM \cite{c1983}, while in 1988 CIPM
recommended a departure from the SI by introducing the practical
electric units \cite{cipm} which have been in effect since 1990.

The desirability of resolving the inconsistency between units used
in precision electric and macroscopic mass measurements and
restoring the SI system as the only system of units for precision
macroscopic measurements drives us to a possible redefinition of
the kilogram and the ampere at the same time. We note that the gap
appeared because the requirement for performing the most precise
electric and mass measurements in the SI units was partly
inconsistent. It still is and may remain for an uncertain period
of time.

The present version of the SI is based on the kilogram prototype
and a fixed value of the magnetic constant $\mu_0$, while the
practical units are based on fixed values of the von Klitzing
constant $R_K$ and the Josephson constant $K_J$ (see
Table~\ref{table} for the values). If we intend to define a
version which allows the derivation of fixed values of $R_K=h/e^2$
and $K_J=2e/h$, we have to fix values of two fundamental
constants, e.g., the Planck constant $h$ and the elementary charge
$e$. To fix two values, we must redefine two units at the same
time.

Thus, the necessary requirement for the redefinition to resolve
the inconsistency is to redefine {\em two\/} units by fixing
values of two fundamental constants at the same time. The
redefinition of the kilogram alone would be of a reduced
importance.

\begin{table}[tbp]
\caption{Numerical values of the involved fundamental constants.
Here, $u_r$ is a relative standard uncertainty. We used units
derived from ${\rm \Omega}_{90}$ and ${\rm V}_{90}$, such as ${\rm
J}_{90}={\rm V}_{90}^2\,{\rm s}/{\rm \Omega}_{90}$, ${\rm
kg}_{90}={\rm J}_{90}\,{\rm s}^2/{\rm m}^2$ etc. The references
are \cite{codata} for CODATA and \cite{cipm} for CIPM. The exact
SI values are from \cite{SI}. The values marked with the asterisk
(*) are derived from related references.} \label{table}
\bigskip
\begin{center}
\begin{tabular}{clccc}
Constant&Value&Unit&$u_r$&Comment\\
\hline
$m({\cal K})$&1 &${\rm kg}$& exactly& SI\\
&$1-1.0(17)\cdot10^{-7}$&${\rm kg}_{90}$& $[1.7\cdot10^{-7}]$&CODATA$^*$\\
$c$&$299\,792\,458$ &${\rm m}/{\rm s}$& exactly&SI\\
$\mu_0$&$4\pi \cdot 10^{-7}$&${\rm N}/{\rm A}^2$& exactly&SI\\
&$4\pi \cdot 10^{-7}\times(1-17.4(3.3)\cdot10^{-9})$&${\rm N}_{90}/{\rm A}_{90}^2$&$[3.3\cdot10^{-9}]$&CODATA$^*$\\
$e$&$1.602\,176\,53(14)\cdot10^{-19}$&${\rm C}$& $[1.7\cdot10^{-7}]$&CODATA\\
&$1.602\,176\,49(66)\cdot10^{-19}$&${\rm C}$& $[4.1\cdot10^{-7}]$&CIPM$^*$\\
&$1.602\,176\,492\dots\cdot10^{-19}$&${\rm C}_{90}$& exactly&CIPM\\
$h$&$6.626\,069\,3(11)\cdot10^{-34}$&${\rm J}\,{\rm s}$&$[1.7\cdot10^{-7}]$&CODATA\\
&$6.626\,068\,9(38)\cdot10^{-34}$&${\rm J}\,{\rm s}$&$[5.7\cdot10^{-7}]$&CIPM$^*$\\
&$6.626\,068\,854\dots\cdot10^{-34}$&${\rm J}_{90}\,{\rm s}$&exactly&CIPM\\
$R_K$ &$25\,812.807\,449(86)$&${\rm \Omega}$& $[3.3\cdot10^{-9}]$&CODATA\\
&$25\,812.807\,0(25)$&${\rm \Omega}$&$[1\cdot10^{-7}]$ &CIPM\\
&$25\,812.807$&${\rm \Omega}_{90}$& exactly& CIPM\\
$K_J$&$483\,597.879(41)\cdot10^{9}$&${\rm Hz}/{\rm V}$&$[8.5\cdot10^{-8}]$ &CODATA\\
&$483\,597.9(2)\cdot10^{9}$&${\rm Hz}/{\rm V}$&$[4\cdot10^{-7}]$&CIPM\\
&$483\,597.9\cdot10^{9}$ &${\rm Hz}/{\rm V}_{90}$&exactly&CIPM\\
$N_A$&$6.022\,141\,5(10)\cdot10^{23}$&$1/$mol&$[1.7\cdot10^{-7}]$&CODATA\\
$h\, N_A$ &$3.990\,312\,716(27)\times10^{-10}$&$\mbox{J\,s/mol}$&$[6.7\times10^{-9}]$&CODATA\\
\end{tabular}
\end{center}
\end{table}

Most of the fundamental constants are related to microscopic
physics (atomic, nuclear or particle physics) and their numerical
values are of two kinds being a result of
\begin{itemize}
\item a pure microscopic comparison (e.g., a value of $m_e/m_p$);
\item a comparison between microscopic and macroscopic values
(e.g., a value of the electron mass in kilograms or eV/c$^2$)
\end{itemize}
The pure microscopic data are more accurate than the data which
involve also macroscopic physics and that is mainly a consequence
of the limited accuracy of measurements linking macroscopic and
atomic physics. Very few numerical values, such as for the
gravitation constant $G$, comes from pure macroscopic experiments,
and these play only a marginal role in precision measurements.

Apparently, without any new experiments we cannot improve the
links between the microscopic and macroscopic physics. However,
the numerical values of the fundamental constants play an
important role as anchor reference data. For instance, it is
customary to express results for X-ray transitions rather in units
of energy (eV) than in terms of the frequency or wave length. To
interpret the frequency as an energy (in eV), one has to apply a
value of $h/e$. We note that the accuracy of comparisons of two
transitions is higher than that of the available numerical value
of $h/e$ in the SI units \cite{codata}. By changing the basis of
the definition of the SI units of mass and charge we can improve
quality of the reference data, and the characterization of the
X-ray transition in terms of electron-volts would be adequate.
This could be achieved by defining the units of mass and charge,
which are now macroscopic, in microscopic terms, i.e., in terms of
$h$ and $e$.

\begin{figure*}[thbp]
\begin{center}
\includegraphics[width=0.6\textwidth]{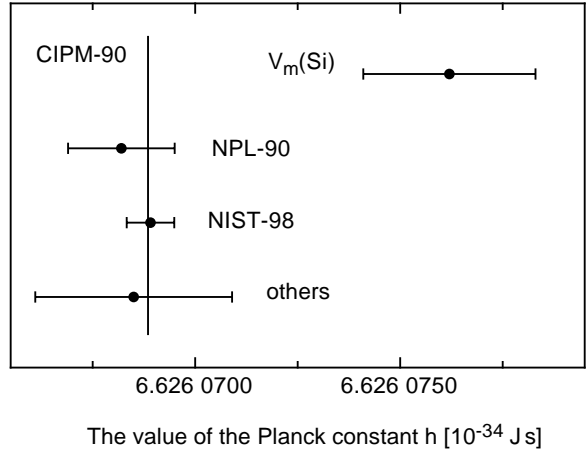}
\end{center}
\caption{Present determinations of the Planck constant $h$. The
watt-balance values (NIST-98 and NPL-90) and XRCD result (V$_{\rm
m}$(Si)) are taken directly from \cite{codata} and labelled in the
same way as there. {\em Others} stands for the average values of
the rest of the data and was communicated to me by Peter Mohr on
base of \cite{codata}. The vertical line indicates a numerical
value of $h$ in practical units \cite{cipm}.} \label{fig}
\end{figure*}

For the SI, the most questionable link between macroscopic and
microscopic physics is related to experiments on determination of
the Planck constant $h$. There is currently an unresolved
discrepancy of 1~ppm between values of the Planck constant derived
from the watt-balance experiments and from the X-ray crystal
density (XRCD) determination (see, e.g., \cite{codata}). The
results of all other measurements together produce a third value
that is competitive in accuracy with the XRCD result and is in a
perfect agreement with the watt-balance values (see
Fig.~\ref{fig}). The importance of this third result is often
underplayed.

These experiments determine a link between the macroscopic mass
unit, the kilogram, and the electric power unit expressed in terms
of volt-90 and ohm-90. This is the crucial link for the
realization of the SI ampere in the present version of the SI. In
the proposed version of the SI \cite{mmqtw}, based on the kilogram
unit defined by a fixed value of the Planck constant $h$ or the
Avogadro constant $N_A$, these experiments determine the mass of
the kilogram prototype.

Recently a number of international commissions and committees
considered this issue. They emphasized the importance of the
problem related to this link and its undesirable effect on
accuracy in mass measurements in the case of the redefinition.
Their concerns are based on an assumption that it is up to those
who decide on the redefinition to involve this link into the SI or
not. We agree that this link is a great problem. But we
unfortunately disagree that this link can be avoided by, e.g.,
postponing the redefinition of the kilogram. This link, as we
mention above, is crucial in present-day realizations of the
ampere (and the volt) of the SI. In other words, it has been used
at least from 1990 for the realization of the SI ampere and there
is no way to avoid this troubled link.

We also raise a question about the conceptual difference between a
constant-based unit and an artefact-based unit. In the latter
case, the definition can have fundamental problems, but it is very
instructive. It is clear in a practical sense what the unit is
and, in the most of comparisons, the method of the comparison is
also obviously fixed. There is not much room for any variety in
realizing the standards. In the former case, when a unit is based
on a constant and certain relations to other quantities (i.e.
certain physical laws), there are a number of ways to realize the
definition and, as in the case of any scientific experiment, the
results may disagree. A substantial difference for a
constant-based unit and an artefact-base unit is due to possible
systematic effects. For the artefact, the systematic effects may
take place, but be reproducible. That is an advantage of the
artefact from a practical point of view. However, such a
systematic effect, being invisible, can produce a drift of the
unit or a reproducible systematic shift (if a way of comparison is
compromised). Various differences in possible realizations of the
constant-based units may produce a discrepancy but that would
allow detection of a possible problem. The very opportunity to
discover the problem, even accompanied by possible discrepancies,
is an advantage.

In the case of a constant-based unit, the systematic effects may
be different. For instance, for determination of the Planck
constant, which is the realization of the SI ampere (presently)
and of the SI kilogram (in the case of the redefinition), these
effects are different and particular results disagree with each
other. Relative mass measurements and relative electric
measurements are more accurate than the link. For the electric
units, CIPM has chosen a clear strategy. A conservative result for
the Planck constant (and, consequently, for the realization of the
SI volt and ampere) has been accepted \cite{cipm}\footnote{We have
to note that any legal adoption of any scientific result (as,
e.g., various {\em mise en pratique}) is an introduction of a
non-SI unit. The SI is a closed system of definitions. Adoption of
anything else as a part of the SI changes the system, while
adoption of anything beyond the SI leads to practical units. In
particular, once the accuracy for the SI values of $K_J$ and $R_K$
is adopted, we arrive at a contradiction between the CIPM
recommendation \cite{cipm} and the accuracy determined by
scientific means \cite{codata}, which is a result of direct
application of the original SI definitions \cite{SI}.}, while the
most accurate measurements are to be performed in practical units.
The same approach should be used for the kilogram in the case of
the redefinition.

In principle, after the redefinition of the kilogram and progress
in its realization, another situation could take place. It may
happen that uncertainty of best realizations and even their
discrepancy (if any) could be smaller than uncertainty related to
the prototype. What strategy should be used to deal with a
possible discrepancy (which in principle could appear from time to
time for any constant-based units)? As long as one particular
method could provide us with reproducible results we should accept
it to define a practical unit, while the related SI unit would be
still defined in a conservative way.

Let us to look now into possible consequences of the redefinition.
First, we need to stress that the only reasonable version of such
a redefinition is to fix $h$ and $e$. We can present certain
advantages in fixing $h$ instead of $N_A$ for the redefinition of
the kilogram only. In particular, the variety, a relatively low
degree of interdependence, and the level of accuracy that has been
achieved makes watt-balance experiment more desirable than the
XRCD measurement. The watt-balance experiment would be a preferred
realization of the kilogram if the Planck constant $h$ is fixed.
On the other hand, the XRCD technique is the most natural choice
for the kilogram based on a fixed value of $N_A$.

One should not overestimate importance of these straightforward
preferences, which are rather educational and practical. However,
they may become of practical importance if the accuracy of mass
measurements increases. As shown in Table~1, the uncertainty of
the molar Planck constant $h\,N_A$ is below 10~ppb. If we fix one
of these two constants, this value would be the uncertainty of the
other. At the present level of accuracy in maintaining the
kilogram, in mass measurements and in the link between the
kilogram and the electric units, this uncertainty is as good as
zero. The accuracy of the link (i.e. of the present-day
determinations of $h$ and $N_A$ separately) for a number of
experimental reasons should be substantially lower than that of
the $h\,N_A$ for a long but uncertain period of time. For a
redefinition of the kilogram alone, it does not much matter which
of these two constants to fix.

However, we have to redefine the kilogram and the ampere at the
same time.  We should clearly choose a redefinition that will
produce fixed values of $R_K$ and $K_J$ to preserve the advantages
of using the present-day practical units. There is no way to do
that unless we fix $h$ and $e$, and we consider this scenario
below. In contrast, if the values of $N_A$ and $e$ are fixed
instead, the electrical community would still need practical units
of resistance and potential. The accuracy of the SI measurements
of those quantities would increase in comparison with the present
situation, but still would be below the one accessible in terms of
the practical units.

If the value of $h$ and $e$ were fixed, then the uncertainties of
the Avogadro constant, the mass of the electron, proton, and
various atoms and nuclei would be substantially smaller. The
uncertainties of the Rydberg constant and various other
frequencies expressed in eV would be greatly improved as well. The
uncertainties of the many electrical measurements in SI units
would be obviously substantially smaller.

Unfortunately, the proposal would have an undesirable effect on
the macroscopic mass measurements in SI units. The consequences of
such effects should be prevented by using the existing prototype
of the kilogram in a way similar to the way the quantum Hall and
Josephson standards are used now. Measurements in terms of the
unit determined by the existing prototype would be considered
measurements in conventional units recommended by CIPM for a
transition period. Macroscopic mass measurements in SI units could
be performed by comparison to the existing prototype together with
its calibration by the measurements which now serve as
determinations of the Planck constant.

How large could the undesirable effects be? At one time it was
assumed that a 10-ppb uncertainty in experiments relating
macroscopic and microscopic masses would be desirable in order to
implement a redefinition of the kilogram. However, it is now felt
that this is not realistic or necessary, particularly in view of
the instability of the mass of the kilogram. In fact, its mass
changed by more than 60~ppb  when it was last washed in the
verification in 1988--1992 (see, e.g., \cite{mmqtw}). However,
this level of accuracy is needed only for experiments like
determination of $h$ and $N_A$. Accuracy, needed for practical
applications is at the level of few parts in $10^7$. Indeed, we
should like to have more accurate standards to perform various
tests, but such tests on, e.g., consistency of various national
standards, may be performed in practical units.

Since the current definition of the SI ampere was essentially
abandoned by the electrical community when conventional values of
the von Klitzing and Josephson constants were introduced in 1988,
there would be no discontinuity in the electrical units if the
ampere were redefined to make the elementary charge $e$ exact and
the kilogram were defined via a fixed value of $h$. In fact, there
would be a significant gain in precision of SI electrical units,
because the SI ohm and volt would be exactly defined in terms of
the quantum Hall and Josephson effects. The present CIPM
recommendations \cite{cipm} on the ohm-90 and volt-90 set the
uncertainty at the level of one and four parts in $10^7$,
respectively, while the most accurate measurements in practical
units are done at the level of few parts in $10^9$ and such
measurements are of practical interest.

The accuracy, continuity, and stability of the units and of the
fundamental constants is critical to the scientific community. In
principle the requirement for the continuity of various units and
constants may be controversial. In particular, the suggested
units, obtained by fixing $h$ and $e$, will not necessarily be the
same as conventional units, since corrections to the standard
expressions for $R_K$ and $K_J$ in terms $h$ and $e$ of are
possible.

There are basically two options.
\begin{itemize}
\item We can set new ohm and new volt to be the same as volt-90
and ohm-90 (except for the necessity of rounding the values in the
definitions). In this case all results in practical units will be
accepted as the SI results, while value of the kilogram and
certain fundamental constants will jump (numerical values of some
constants in the SI units and units-90 are presented in
Table~\ref{table}).
\item We can choose an option to adopt values of $h$ and $e$ as
they are in the CODATA paper \cite{codata} (or the newest
available CODATA results at the time of the redefinition). That
will reduce a possible jump\footnote{The very existence of the
jump becomes questionable and some believe that there would be no
jump at all. We choose here to use a common word `jump' instead of
`discontinuity' because in a sense there would be no
discontinuity, but there should be a certain jump. The jump would
be a result of a two-step action. For instance, we use an exact
value of $\mu_0$ \cite{SI}. With the redefinition, we make it
measurable. Once we use the CODATA data (see Table~\ref{table}),
the new result for $\mu_0=2\alpha R_K/c$ should have an
uncertainty. The new result may be consistent with the presently
fixed value of $\mu_0$ \cite{SI}. Nevertheless, sooner or later
with improvement of accuracy, a value of $\mu_0$ would depart from
the previously fixed numerical value. There is no chance that we
can guess the values for $e$ and $h$ in such a way that the
measurable $\mu_0$ would be exactly the same as before. That is
the same as trying to guess an exact value of the fine structure
constant $\alpha$. So, eventually a certain jump would take place,
but in each step we should have no discontinuity.} in the kilogram
of SI and values of the fundamental constants.
\end{itemize}

Technically we can fix first $R_K$ and $K_J$, calculate $h$ and
$e$, and round them properly at the end. It is more transparent to
discuss consequences of fixing different values of $R_K$ and $K_J$
than of $h$ and $e$.

At present, CODATA's $\{R_K\}_{\rm SI}$ differs from CIPM's
$\{R_K\}_{\rm 90}$ by approximately five standard deviations (see
Table~\ref{table}) and we have to choose between them. On the
other hand, CODATA's $\{K_J\}_{\rm SI}$ differs from CIPM's
$\{K_J\}_{\rm 90}$ by less than one standard deviations (see
Fig.~\ref{fig}) and we can choose either of them without any
serious consequences.

A choice between different values of $R_K$ would affect a value of
$\mu_0$ and a possible departure of the new SI ohm from the
present SI ohm. If we choose the CIPM's value, the new ohm SI
would be related to ohm-90 and to the present CIPM central value
of the SI ohm (we cannot discuss here the SI ampere because its
definition involve $\mu_0$ and the kilogram, which is also a
subject of changes). A value of $\mu_0$ would depart from its
present value, but it is not particularly important for precision
measurements. The only experiments are with calculable capacitors,
but there are very few of them around the world and a shift at the
level below 20~ppb is not important at their present level of
realization.

In principle, impact of choice between different values of $K_J$
could be more important. Different $K_J$ would lead to different
values of the kilogram, the volt and the ampere and different
numerical values of various important fundamental constants such
as $h$, $e$, particle masses in kilograms and eV/c$^2$, and energy
of various atomic and nuclear transitions in electron volts.
Fortunately, as we mention above, CODATA's $\{K_J\}_{\rm SI}$ and
CIPM's $\{K_J\}_{\rm 90}$ agree to each other within a standard
deviation (the difference is approximately half a deviation) and
perhaps, because of this agreement, we should choose the CIPM's
value. A change in a value of the fundamental constants within one
sigma is not a discontinuity, and we would prefer to set the
redefined SI units to be equal to the practical units.

That does not downplay the importance of the CODATA values. The
CODATA evaluation will determine a recommended value of the
magnetic constant $\mu_0$ and a value of the mass of the prototype
$m({\cal K})$. The situation with $K_J$ may change by 2007 and
with new watt-balance and XRCD results the CODATA's $\{K_J\}_{\rm
SI}$ could depart from $\{K_J\}_{\rm 90}$. If the difference would
be above one standard deviation we will need to make a real choice
between CODATA's and CIPM's values.

We believe that the kilogram and the ampere should be redefined
and that they should be redefined at the same time by fixing
values of the fundamental constants $h$ and $e$. Two open
practical questions are related to choice for the fixed values
(discussed above) and to a proper time for the redefinition.

A choice for the timing should consider the following.
\begin{itemize}
\item Any decision (positive or negative) on the redefinition will
have benefits and expenses. These have to be examined carefully.
\item Some disadvantages (discontinuity in values of units and
numerical values of the constants, worsening of accuracy of
measurements in SI units, etc.) may be unavoidable. It is
necessary to take into account that postponing a necessary
decision could increase expenses.
\end{itemize}

Considering the advantages and disadvantages we point that
\begin{itemize}
\item from the point of view of mass metrology, the redefinition
of the kilogram will be successful once the Planck constant is
reliably determined with a standard uncertainty less than about
50~ppb;
\item from the point of view of electric measurements, the
redefinition will be successful even now because of the immediate
improvement in accuracy of precision electric measurements in SI
units;
\item the final decision on the proper time for the redefinition
of the kilogram and the ampere can be made when there is a net
gain based on a careful comparison of the advantages and
disadvantages. Careful examination should be given to the relative
importance to the fields of electric and mass measurements
(accuracy, volume of measurements, area of applications etc.)
where the changes would take place.
\end{itemize}
The standards themselves have no value if they are not needed for
actual or future applications. For this it is most important for
us is to consider the consequences outside of the standards
community. Metrologists are trained to deal with different units
(the SI units and various practical units), while outside people
are not. There is a limited number of scientific experiments were
such a level of accuracy is important. Improvement of the
reference data is clearly an advantage for outsiders.

We hope that a real study on the practical importance of precision
mass and electric measurements with uncertainty below a 100~ppb
will be done. Up to now, despite numerous considerations of the
redefinition, this question has not been discussed at all, or at
least the results of such a discussion have not been made
available.

This paper is an extended version of document CCU/05-27, a working
document of the 17th meeting of of Consultative Commitee for
Units. During discussions there and at the preceding meeting of
the CODATA task group on fundamental constants, it became clear
for me that the accuracy actually demanded in precision electric
measurements are about two orders of magnitude higher than in mass
measurements. Still, examination of the problem is necessary,
because a question is not only accuracy, but also number of the
measurements.

To conclude this paper let us suggest wording for the redefinition
of the kilogram: ``{\em The kilogram is the mass of a body whose
rest energy is equal to the energy of\/}
$299\,792\,458\cdot10^{27}$ {\em optical photons in vacuum of
wavelength of\/} $66.606\,9311$ {\em nanometres.''\/} The explicit
indication of the number of the photons is necessary\footnote{I
discussed the issue on the number of the photons with Peter Mohr
and it resulted in footnote 2 in \cite{mmqtw}. I need to mention
that I am impressed by elegancy of the solution there of another
problem of the definition, which is avoiding of a multiplication
and division of numerical values of constants $c$ and $h$ in such
a combination as $\{c\}^2/ \{h\}$. The version presented here is
somewhat different from that in \cite{mmqtw}. From a point of view
of relativistic physics it is preferable to speak in terms of the
rest energy rather than the rest mass (see, e.g., \cite{okun}).}.
The kilogram is a macroscopic quantity, while we try to link it to
a microscopic object. With a single microscopic object we arrive
at the situation when the energy is bigger than the Planck energy
or the wave length is shorter than the Planck length.

Concerning the ampere, a redefinition in terms of the elementary
charge is rather trivial. Still, it should be mentioned, that,
when the definition of the ampere was adopted, its direct
realization was possible. Now, we see that there are two units
(the ohm and the volt), which we can realize directly and two
units (the ampere and the coulomb) which are related to more
fundamental quantities but cannot be realized directly. In former
time, the ampere was a good choice. Now, if we like to make a
practical choice it should favors the volt (potential is more
fundamental than resistance), and if we like to make a physical
choice we should prefer the coulomb (in particular, because of its
educational advantages). There are no advantages for the ampere
anymore.

The author is grateful to R. Davis, J. Flowers, P. J. Mohr, L. B.
Okun, L. Pendrill, B. N. Taylor and B. M. Wood for useful and
stimulating discussions.


\begin{thebibliography}{9.}

\bibitem{mmqtw} I. M. Mills, P. J. Mohr, T. J. Quinn, B. N. Taylor and E. R. Williams, Metrologia {\bf 42}, 71 (2005)

\bibitem{SI} {\em The International System of Units (SI)\/}. BIPM, S\`evres, 1998.

\bibitem{c1983} Metrologia {\bf 20}, 25 (1983).

\bibitem{cipm}
T. J. Quinn, Metrologia {\bf 26}, 69 (1989) (see p. 70);\\ T. J.
Quinn, Metrologia {\bf 38}, 89 (2001) (see p. 91);\\ T. J. Quinn,
Metrologia {\bf 26}, 69 (1989) (see p. 69).

\bibitem{codata} P. J. Mohr and B. N. Taylor, Rev. Mod. Phys.
{\bf 77}, 1 (2005).

\bibitem{okun} L. B. Okun Nucl. Phys. Proc. Suppl. {\bf 110} (2002)
151.

\end{thebibliography}
\end{document}